# A Multi-agent Framework for Performance Tuning in Distributed Environment

Sarbani Roy, Saikat Halder and Nandini Mukherjee

Department of Computer Science and Engineering,
Jadavpur University, Kolkata 700 032, India
Email nmukherjee@cse.jdvu.ac.in

**ABSTRACT:** This paper presents the overall design of a multi-agent framework for tuning the performance of an application executing in a distributed environment. The multi-agent framework provides services like resource brokering, analyzing performance monitoring data, local tuning and also rescheduling in case of any performance problem on a specific resource provider. The paper also briefly describes the implementation of some part of the framework. In particular, job migration on the basis of performance monitoring data is particularly highlighted in this paper.

## 1. INTRODUCTION

Computational resources on a Grid together can solve very large problems requiring more resources than is available on a single machine. An application is benefited from a Grid environment when the resource requirement cannot be fulfilled (either quantitatively[1] or qualitatively[2]) from the resources owned by the user. Thus, Grid provides a good basis of creating a collaborative environment for high performance computing. An application running on a Grid needs to be adaptive with its current execution environment so that it can efficiently utilize the available resources. The basic technique is to regularly monitor the infrastructure and the application performance and tune the application by using optimization techniques or adding more resources to its execution environment or simply by migrating the application or its components. In this paper we propose a multi-agent framework for providing adaptability to an application running on a distributed system. The framework proposed here is general enough to cope with any distributed system with some extent of dynamism, although our aim is to implement the framework for a Grid environment.

## 2. A MULTI-AGENT FRAMEWORK FOR ADAPTIVE EXECUTION OF APPLICATIONS

In this section we describe the overall design of a multi-agent framework. Within the framework, an application dynamically adapts to changing resource availability exploiting adaptive runtime systems to negotiate program behavior and resource management. The framework comprises four different components, which work in an integrated manner. These four components are: *a) A resource broker b) A Job Controller c) An analyzer d) A performance tuner.*

The *Resource Broker* acts as an intermediary between the application and a set of resources. It is the responsibility of the *Resource Broker* to negotiate and find suitable resources according to the application's resource requirement. The *Resource Broker* provides a single point of submission for jobs and forwards the jobs to one of the underlying resources for execution using standard resource management protocols (such as GRAM [1] in a Grid environment).

---

[1] to provide more resources than available locally. Thus, resources may be unified to speed the execution or to execute significantly larger and more complex problems.
[2] to provide more special resources than available locally. Performance is linked with the resource quality and availability.

The *Job Controller* is responsible for controlling the execution of the application at the local level. It also maintains a global view of the application's runtime behavior, as well as functioning of the infrastructure and performs control actions for improving the performance whenever the *service-level agreement* (SLA) [4] is violated.

The *Analyzer* component monitors individual resources and gathers performance monitoring data related to the execution behaviour and infrastructure functioning. The application performance analysis data consists of an evaluation of performance properties and identification of regions showing the performance problems [2, 3]. A global part of the *Analyzer* also maintains an integrated view of the monitoring data obtained from individual resources. This helps in identifying any performance bottleneck in the entire system.

The *Performance Tuner* is responsible for tuning performance of the application at local level. It also maintains a global view of the application's runtime behavior, as well as functioning of the infrastructure and performs control actions for improving the performance whenever a *service-level agreement* is violated. Thus, the *Performance Tuner* consists of two different subcomponents, one that acts locally and the other, that acts on a global basis. The one that acts locally works with local policies and interacts with local resource management services. The global component of the performance tuner interacts with the *Resource Broker* and takes steps, such as rescheduling a job or establishing new *service-level agreements*.

A set of agents is responsible for all interactions within and among the components. These agents cooperate with each other to provide an adaptive execution environment for the application. We distinguish the agents in two different categories: (i) functional agents and (ii) control agents. The functional agents in our framework perform specific tasks that are entrusted to them. On the other hand, control agents are responsible for controlling the execution of the application. Each type of agents is part of either the *Resource Broker*, or the Job Controller, or the *Analyzer*, or the *Performance Tuner*.

Our system uses three types of functional agents, which are described below.

**Broker Agent:** The *Broker Agent* resides in the Resource Broker component. The Broker agent receives the Job Requirement List (JRL), where the basic requirements of the job are defined. The Broker Agent consults the Grid Information Services (GRIS/GIIS, MDS of Globus [6]) to obtain information about the available resources and prepare Resource Specification Table (RST) for Resource Providers willing to provide computational service. Where the information about a single resource provider is known as the Resource Specification Template and the collection of several such templates is known as Resource Specification Table (RST). B*roker Agent* matches the JRL with the RSTs and finds suitable resources for a particular job.

**Analysis Agent:** *Analysis agent* resides in the *Analyzer* and evaluates performance properties and detects performance problems. In addition to the resource-based analysis agents, there are agents that monitor the overall performance of the Grid. Thus an agent hierarchy is formed in which the lowest level agents monitor individual resources and the higher level agents collect data from the local agents and analyses the data in order to detect any performance bottleneck in the system.

**Tuning Agent:** *Tuning agent* resides in Performance Tuner, and responsible for local tuning of the running jobs on a specific resource provider.

Two types of *Control Agents,* part of the Job Controller, control the execution of an application. One is designated as supervisory type and it globally looks after the execution of all parallel jobs of an application. It is responsible for taking all global decisions, including rescheduling and establishing new SLAs [4]. The *Supervisory Control Agent* maintains a list of resource provider addresses for each job for future rescheduling.

Other control agents are subordinate to the supervisory agents. Each *Subordinate Control Agent* is associated with one of the parallel jobs of the application. It acquires the job along with an SLA from the *Supervisory Control Agent* and gets deployed on the resource provider by a *scheduler*. The agent then resides on the resource provider. At the time of rescheduling, the subordinate agent carries the job to the new resource provider. The *Subordinate Control Agents* are lightweight, mobile agents that, under the directives of the supervisory agent, independently carry the jobs to the resource providers, submit the jobs and monitor their execution.

## 3. IMPLEMENTATION

This section discusses in detail the preliminary implementation of the ideas presented above and the initial results. Using our multi-agent framework we developed a fault tolerant, adaptive runtime infrastructure that incorporates both performance evaluation and the persistent checkpointable computational environment for a cluster of workstations. We use PAPI [5] to monitor the runtime behaviors of any on-time executing job. A Java interface has been built as an upper layer on top of

Performance APIs of PAPI to access different cross platform hardware preset or native events. The *Analysis Agent* sits on top of PAPI and the Java interface and detects any performance problem on the basis of PAPI-generated monitoring data.

To provide the fault tolerant aspect to any application running on a local node we need to checkpoint it periodically and upon emergency migrate it to a distant node for later part of the execution. The total execution of the job is thus being carried out by different JVMs at different times, if necessary, before transporting the result back to the initiating local JVM. Our mobile *Subordinate Agents* are assigned with the responsibility of migration of the jobs on the basis of the report sent by the *Analysis Agents*. The persistent fault tolerant model used in the current work incorporates an efficient incremental checkpointing strategy for any running application and migrate its suspended state to any remote JVM for restoring execution.

The checkpointing and recovery concepts have been prototypically implemented. The prototype provides generic mechanisms for extracting the executing object persistent state and stores the suspended state on a persistent store before deciding to transport to a distant idle node with distinct JVM for the remaining execution. The fault tolerant application layer implements the incremental checkpointing and monitoring process, while the *Subordinate Agents* within the Mobile Agent Framework serve the purpose of relocating the suspended job to a different site. Checkpointing (making a copy of the state of an object) is done using Java serialization techniques. This creates a copy of the object, meaning all member variables are themselves copied rather than just their references. The design layers as implemented in the current version and the migration of Checkpointed job are presented in Figure 1 (a) and (b) respectively.

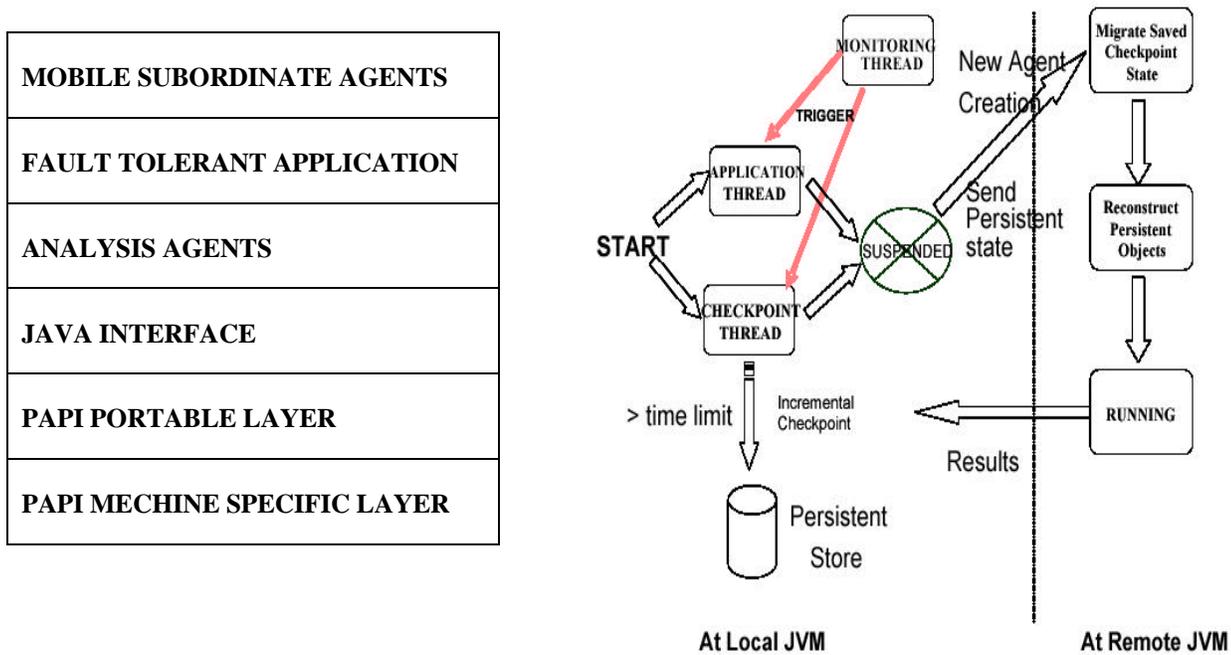

Figure 1: (a) Different Design Layers.                          (b) Migration of Checkpointed Job.

We tested our preliminary implementation on a system comprising a node having an Intel Pentium-4 processor of 2.8 GHz speed and 512 MB physical memory as server1 and a node with Intel Xeon processor of 3.0 GHz speed and 1024 MB physical memory as server2. Table 1 depicts the effect of rescheduling, by comparing the outcomes of the following two scenarios.

**Scenario1:** The Subordinate agent is initially scheduled to server1 and completes its execution on it.

**Scenario2:** The Subordinate agent is initially scheduled to server1, and at a certain instance of time it is rescheduled to server2 (because the server1 withdraws its services) and executes the remaining part of the job on server2.

An overhead is involved in the second case because of rescheduling. The total time required to complete the execution in both cases as well as the time spent on each server in the second case and the overhead for rescheduling are shown in Table 1. Figure 2 (a) compares the total time taken by the subordinate agent for Scenario1 and Scenario2. It is clear that rescheduling is good choice if server1 withdraws its services. Although rescheduling from server1 to server2 does not achieve significant improvement in running time (possibly due to little difference in the server configurations), it however demonstrates that rescheduling can be done successfully through mobile agents. We could obtain improved results if we have chosen an enhanced system as a server2, such as a multi-processor system. The test code is a simple sorting program. We varied the size of the array to be sorted and measured the execution time on server1 for Scenario1 and execution times on server1 and server2 for Scenario2.

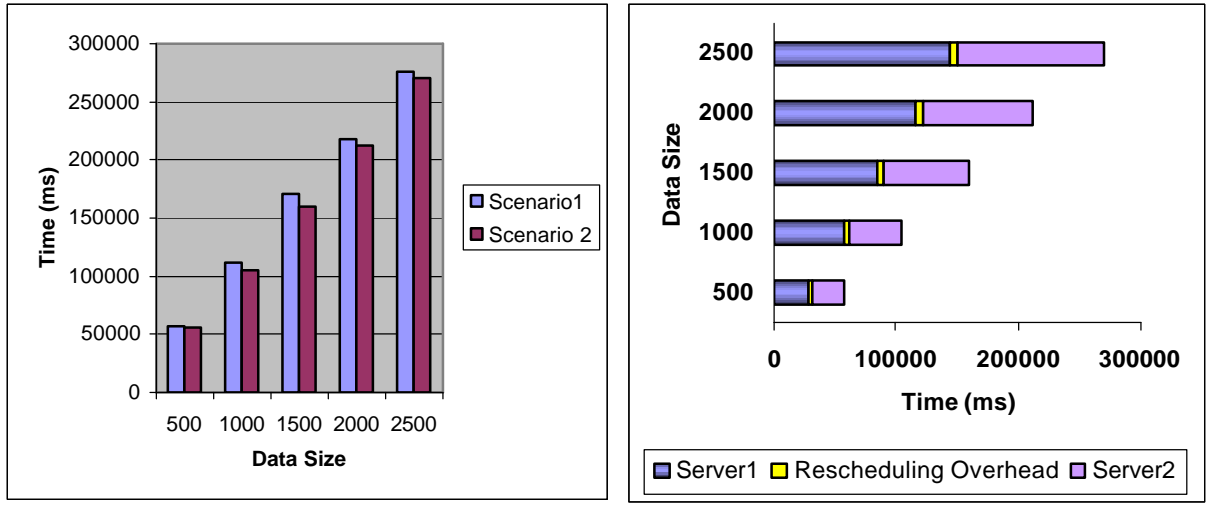

Figure2: (a) Comparison of the performances in Scenario 1 and Scenario 2  (b) Results of Rescheduling from Server1 to Server2

| Data Size N | Scenario1<br>Total time on Server1 (ms) | Scenario2<br>Rescheduling Total time (ms) | Number of iterations before migration from Server1 to Server2 | Time spent on Server1 before migration (ms) | Time spent on Server2 after migration (ms) | Overhead (ms) |
|---|---|---|---|---|---|---|
| 500 | 57306 | 56422 | 249 | 27381 | 25421 | 3620 |
| 1000 | 111686 | 104896 | 516 | 56805 | 43371 | 4720 |
| 1500 | 171436 | 159883 | 764 | 84056 | 70002 | 5825 |
| 2000 | 217751 | 212264 | 1050 | 115500 | 89811 | 6953 |
| 2500 | 276016 | 270604 | 1298 | 142882 | 119944 | 7778 |

Table 1: Execution Performance of Subordinate agent for simple sorting of N data, initially scheduled to Server1 (P-IV) and after i iterations migrated to Server2 (Xeon), times are in milliseconds.

## 4. RELATED WORK

One of the major objectives of research in Performance Engineering is maintaining performance QoS for individual applications. The ICENI project emphasises a component framework for generality and simplicity of use [6]. Application performance is achieved by an application mapper, which selects the "best" component implementations for the resources available, based on component meta-data. The GrADS project focuses on building a framework for both preparing and executing applications in Grid environment [7]. Each application has an application manager, which monitors the performance of that application for QoS achievement. Failure to achieve QoS contract causes a rescheduling or redistribution of resources. Both ICENI and GrADS have resource schedulers which partition resources between applications. GrADS monitor resources using NWS and use Autopilot for performance prediction [8, 9]. A performance steering system developed within the RealityGrid project [10] has been described in [11]. The aim is to allow scientists to explore component-based simulations of physical phenomena, while simultaneously steering the performance of the simulations in an autonomous manner. The system, unlike GrADS and ICENI, is only concerned with single applications. The framework proposed in this paper works with similar idea. But unlike the previous systems, it employs autonomous agents for carrying out the performance tuning jobs.

## 5. CONCLUSION AND FUTURE WORK

The paper presents a multiagent framework for performance tuning of applications in Distributed environment. Our framework provides an adaptive runtime environment for execution of applications. In future, the framework will be implemented on top of the available Grid middleware (such as Globus Toolkit [1]), and will use the services available with it.

A simple version of the Analysis Agent has so far been implemented. This will be further modified to capture all types of performance problems and evaluate the performance properties of the infrastructure and the application. We have not yet implemented the Supervisory Control Agent. In our next course of action, the Supervisory Agent will be implemented so that it can keep track of all the parallel jobs of an application running in the distributed environment. We are also going to implement our framework using jini [12] to investigate its conveniences in distributed environment.

We already have tested simple applications like sorting. However, more complex applications requiring intensive computational services need to be tested within our framework. Also the current implementation of checkpointing and job migration has several performance issues that need to be addressed in our future implementations.